\documentclass[galaxies,article,accept,moreauthors,10pt,a4paper]{mdpi}     
\firstpage{1} 
\articlenumber{x}
\doinum{10.3390/galaxies------}
\pubvolume{xx}
\pubyear{2017}
\copyrightyear{2017}
\externaleditor{Academic Editor: Emmanouil Angelakis, Markus Boettcher
Jose L. G\`omez}
\history{Received: date; Accepted: date; Published: date}

\usepackage{microtype}
\usepackage{soul}
\usepackage{subfigure}

\Title{Microscopic Processes in Global Relativistic Jets Containing Helical Magnetic Fields: Dependence on Jet Radius
}

\orcidauthorONE{0000-0001-6031-7040} 

\Author{{Ken-Ichi Nishikawa} $^{1,}$*, Yosuke Mizuno $^{2}$,   Jose L. G\'{o}mez $^{3}$, Ioana Du\c{t}an $^{4}$, Athina Meli $^{5,6}$, Charley White $^{7}$, 
Jacek Niemiec $^{8}$, Oleh Kobzar $^{8}$,
 Martin Pohl $^{9,10}$, Asaf Pe'er $^{11}$,
 Jacob Trier Frederiksen $^{12}$,  \AA ke Nordlund $^{13}$,  Helene Sol $^{14}$,
 Philip E. Hardee $^{15}$
 \mbox{and Dieter H. Hartmann $^{16}$ }}

\address{%
$^{1}$ \quad  Department of Physics, University of Alabama in Huntsville, ZP12, Huntsville, AL 35899, USA\\
$^{2}$ \quad Institute for Theoretical Physics, Goethe University,  Frankfurt am Main {D-60438}, Germany; mizuno@th.physik.uni-frankfurt.de\\
$^{3}$ \quad Instituto de Astrof\'{i}sica de Andaluc\'{i}a, CSIC, Apartado 3004, Granada 18080, Spain; jlgomez@iaa.csic.es\\
$^{4}$ \quad Institute of Space Science, Atomistilor 409, Bucharest-Magurele RO-077125, Romania; ioana.dutan@gmail.com\\
$^{5}$\quad Department  of Physics and Astronomy, University of Gent, Proeftuinstraat 86,
Gent B-9000, Belgium\\
$^{6}$\quad Department  of Physics and Astronomy, University of Liege,
Place du 20-Ao\^ut, 7 4000 Li\`ege, Belgium; ameli@ulg.ac.be\\
$^{7}$ \quad  Department of Physics, University of Alabama in Huntsville,  Huntsville, AL 35899, USA; cmw0037@uah.edu\\
$^{8}$ \quad Institute of Nuclear Physics PAN, ul. Radzikowskiego
152,  Krak\'{o}w 31-342, Poland; Jacek.Niemiec@ifj.edu.pl~(J.N.); oleh.kobzar@ifj.edu.pl{(O.K.)}\\
$^{9}$ \quad  Institut fur Physik und Astronomie, Universit\"{a}t Potsdam,  Potsdam-Golm {14476}, Germany; pohlmadq@gmail.com\\
$^{10}$ \quad DESY, Platanenallee 6,  Zeuthen {15738}, Germany; pohlmadq@gmail.com\\
$^{11}$ \quad Physics Department, University College Cork, Cork T12 YN60, Ireland; a.peer@ucc.ie\\
$^{12}$ \quad Innofactor Denmark A$/$S, Telia Parken, \O ster All\'e 48,
2100 K\o benhavn \O, Denmark; jacob.trier@innofactor.com\\
$^{13}$ \quad Niels Bohr Institute, University of Copenhagen, Blegdamsvej 17, K\o benhavn DK-2100, Denmark;  aake@nbi.dk\\

$^{14}$\quad LUTH, Observatore de Paris-Meudon, 5 place Jules
Jansen, Meudon Cedex 92195, France; helene.sol@obspm.fr\\
$^{15}$\quad Department of Physics and Astronomy, The University
of Alabama, Tuscaloosa, AL 35487, USA; pehardee@gmail.com\\
$^{16}$\quad Department of Physics and Astronomy, Clemson University, Clemson, SC 29634, USA; hdieter@g.clemson.edu}

\corres{Correspondence: ken-ichi.nishikawa@nasa.gov; Tel.: +1-256-824-2593}

\firstnote{Current address: Affiliation 7:  Department of Physics Florida State University 
600 W. College Avenue Tallahassee, FL 32306, USA} 
\secondnote{These authors contributed equally to this work.}

\abstract{In this study we investigate jet interaction at a microscopic level in a cosmological environment, which responds to a key open question in the study of 
relativistic jets. Using small simulation systems during prior research, we initially studied the evolution of both electron-proton and electron-positron relativistic jets containing helical magnetic fields, by focusing on their interactions with an ambient plasma.
Here, using larger jet radii, we have performed simulations of  global jets containing helical magnetic fields in order to examine how helical magnetic fields affect kinetic instabilities  such as the Weibel instability, the kinetic Kelvin-Helmholtz instability (kKHI)  and the Mushroom instability (MI). 
We  found that the evolution of global jets strongly depends on the size of the jet radius. For example, phase bunching of jet electrons, in particular in the 
electron-proton jet, is mixed with larger jet radius due to the more complicated structures of magnetic fields with excited  kinetic instabilities.
In our simulation study these kinetic instabilities lead to new types of instabilities in global jets.  In the electron-proton jet simulation  a modified recollimation occurs and jet electrons are strongly perturbed. In the electron-positron jet simulation mixed kinetic instabilities occur at early times followed by a turbulence-like structure. Simulations using much larger (and longer) systems are further required in order to thoroughly investigate the evolution of global jets containing helical magnetic fields.}

\keyword{relativistic jets; particle-in-cell simulations; global jets; helical magnetic fields; kinetic instabilities;
kink-like instability; recollimation shocks; polarized radiation} 

\begin{document}


\section{Introduction}

Relativistic jets are collimated plasma outflows associated with active galactic nuclei (AGN), gamma-ray bursts 
(GRBs), and pulsars (e.g., \cite{haw15}). Among these astrophysical systems, blazars and GRB jets produce the most luminous phenomena 
in the universe (e.g., \cite{peer14}). Despite extensive observational and theoretical investigations, including 
simulation studies, our understanding of their formation, their interaction and evolution in the ambient plasma, 
and consequently their observable properties such as  time-dependent flux and polarity~{(}e.g., \cite{mm16}{)}, remains quite limited. 

Astrophysical jets are ubiquitous in the universe and provide many essential plasma phenomena such as interaction with interstellar medium,  generation
of magnetic  fields, turbulence, reconnection and particle acceleration. Many of the processes that determine the behavior of global relativistic jets are very 
complex, involving plasma physics and often coupling global, large-scale dynamics to microscopic processes that occur on short spatial and temporal 
scales associated with plasma kinetic effects. We have carried out  kinetic plasma simulations using our relativistic particle-in-cell (RPIC) code, with the intent
to advance our knowledge of global relativistic jets with  helical magnetic fields and associated phenomena such as particle acceleration, 
kinetic reconnection and turbulence, which cannot be investigated with fluid models (i.e., relativistic magnetohydrodynamic (RMHD) simulations).

Recently, we performed ``global'' jet simulations involving {the} injection of a~cylindrical unmagnetized jet into an ambient plasma in order to investigate shock (Weibel instability) and velocity shear instabilities (kKHI and {Mushroom instability (MI))} simultaneously \cite{nishi16a}. 
In this paper we will describe the preliminary results of this new study of global relativistic jets containing helical magnetic fields.

 
\section{Simulation Setups of Global Jet Simulations} 

Jets generated  from black holes, which are then  injected into the ambient interstellar medium contain magnetic fields that are thought to be helically twisted. 
Based on this understanding, we performed global simulations of jets containing helical magnetic fields injected
into an ambient medium
~(e.g., \cite{nishi16b,dutan17}).
The key issue we investigated is how the helical magnetic fields affect the growth of the kKHI, the MI, and the Weibel instability.  RMHD
~simulations demonstrated that jets containing helical magnetic fields develop 
~{kink} instability
~{(}e.g.,~\cite{mizuno14,singh16,barn17}{)}.  Since our RPIC simulations are large enough to include a kink instability, we found a kink-like instability in our
electron-proton jet case.  

\vspace*{-0.0cm}
\subsection{Helical Magnetic Field Structure}
\vspace*{-0.0cm}

In our simulations \cite{nishi16b}, cylindrical jets are injected with a helical magnetic field 
implemented like that in RMHD simulations performed by Mizuno et al. \cite{mizuno15}. 
Our simulations use Cartesian coordinates.
Since $\alpha =1$, Equations (9)--(11) from \cite{mizuno15} are reduced to Equation (1), and the magnetic field  takes the form:
\begin{eqnarray}
B_{x} = \frac{B_{0}}{[1 + (r/a)^2]}, \,  \, \, \, \, \,  B_{\phi} =  \frac{(r/a)B_{0}}{[1 + (r/a)^2]}
\end{eqnarray}

The toroidal magnetic field is created by a current $+J_{x}(y, z)$ in the positive $x$-direction, so that
defined in Cartesian coordinates:
\begin{eqnarray}
B_{y}(y, z) =  \frac{((z-z_{\rm jc})/a)B_{0}}{[1 + (r/a)^2]}, \, \, \,\,  \,  \,
B_{z}(y, z) =  -\frac{((y-y_{\rm jc})/a)B_{0}}{[1 + (r/a)^2]}.
\end{eqnarray}

Here  $a$ is  the characteristic  length-scale of the helical magnetic field, $(y_{\rm jc},\, z_{\rm jc})$ is the center of
the jet, and  $r = \sqrt{(y-y_{\rm jc})^2+(z-z_{\rm jc})^2}$. The chosen helicity is defined through Equation (2), which
has a left-handed polarity with positive $B_0$. At the jet orifice, we  implement  the helical magnetic field without the motional electric fields. This corresponds to a toroidal magnetic field generated self-consistently by jet particles moving along{ the} $+x$-direction.


\vspace{-0.0cm}
\subsection{Magnetic Fields in Helically Magnetized RPIC Jets with Larger Jet Radius}
\vspace{-0.0cm}

As an initial step, we  examined how the helical magnetic field modifies jet evolution using a~small system before performing larger-scale simulations. 
A schematic of the simulation injection setup  is used in our previous work \cite{nishi16b}. In these small system simulations, we utilized a numerical grid with 
\mbox{$(L_{x}, L_{y}, L_{z}) = (645\Delta, 131\Delta, 131\Delta)$} (simulation cell size: $\Delta = 1$) and periodic boundary conditions in transverse directions with 
jet radius $r_{\rm jt} = 20\Delta$. 
The jet and ambient (electron) plasma number  density measured in the simulation 
 frame is $n_{\rm jt}= 8$ and  $n_{\rm am} = 12$,  respectively. This set of densities of jet and ambient plasmas is used in our previous simulations \cite{nishi16a,nishi16b,dutan17}
 
 In the simulations, the electron skin depth  $\lambda_{\rm s} =  c/\omega_{\rm pe} = 10.0\Delta$, where  $c$ is the speed of light,  
$\omega_{\rm pe} = (e^{2}n_{\rm am}
/\epsilon_0 m_{\rm e})^{1/2}$ is the electron plasma frequency, and  the electron Debye length for the ambient
electrons  {is} $\lambda_{\rm D}=0.5\Delta$.  The jet--electron thermal velocity is $v_{\rm jt,th,e} = 0.014c$
in the jet reference frame.  The electron thermal velocity in the ambient plasma is
$v_{\rm am,th,e} = 0.03c$, and  ion thermal velocities are smaller by $(m_{\rm i}/m_{\rm e})^{1/2}$.
Simulations were performed using an electron--positron ($e^{\pm}$) plasma or
an electron--proton ($e^{-}$ {--}$p^{+}$
~with $m_{\rm p}/m_{\rm e} = 1836$) plasma for the jet Lorentz factor of 15
and with the ambient plasma at rest  ($v_{\rm am}= 0$).

In the simulations  we use the initial magnetic field amplitude parameter $B_{0}=0.1c$,  {(}$c=1$ {)},
~\mbox{ {(}$\sigma = B^2/n_{\rm e}m_{\rm e}\gamma_{\rm jet}c^{2} =2.8\times 10^{-3}$ {)}}, and $a =  0.25*r_{\rm jt}$. The helical field structure inside the jet is defined by Equations (1) and (2). For the external magnetic fields, we use a damping function $\exp{[-(r-r_{\rm jt})^{2}/b]}$ $(r \ge r_{\rm jt})$ that multiplies Equations (1) and (2) with the tapering parameter $b=200$. The final profiles of the helical magnetic field components are similar to that in
the case where jet radius $r_{\rm jt} = 20\Delta$, the only difference is $a= 0.25*r_{\rm jt}$. 

%
%

In this report we maintain all simulation parameters as described above except jet radius and simulation size (adjusted based on the jet radius). 
We have performed simulations 
with  larger jet radii $r_{\rm jt} =40\Delta,\, 80\Delta,\, {\rm and}\, 120\Delta$.  
 In these small system simulations, we utilize a numerical grid with \mbox{$(L_{x}, L_{y}, L_{z}) = (645\Delta, 257\Delta, 257\Delta), (645\Delta, 509\Delta, 509\Delta), (645\Delta, 761\Delta, 761\Delta)$} (simulation cell size: 
 $\Delta = 1$).   The cylindrical jet with jet radius $r_{\rm jt} =40\Delta,\,  80\Delta,\, {\rm and}\, 120\Delta$ is  injected in the middle of   
 {the} $y$ -- $z$
~plane ($(y_{\rm jc}, z_{\rm jc}) = (129\Delta, 129\Delta), (252\Delta, 252\Delta), (381\Delta, 381\Delta)$) at $x= 100\Delta$, respectively.
The largest jet radius ($r_{\rm jt} =120\Delta$) is larger than that ($r_{\rm jt} =100\Delta$) in \cite{nishi16a}, but the simulation length is much shorter ($_{x} = 2005\Delta$).

\begin{figure}[htbp]
\vspace*{-0.3cm}
\hspace{7.20cm} (a) \hspace{7.0cm} (b)

\vspace{-0.cm}

\hspace{0.3cm}
\includegraphics[scale=0.43]{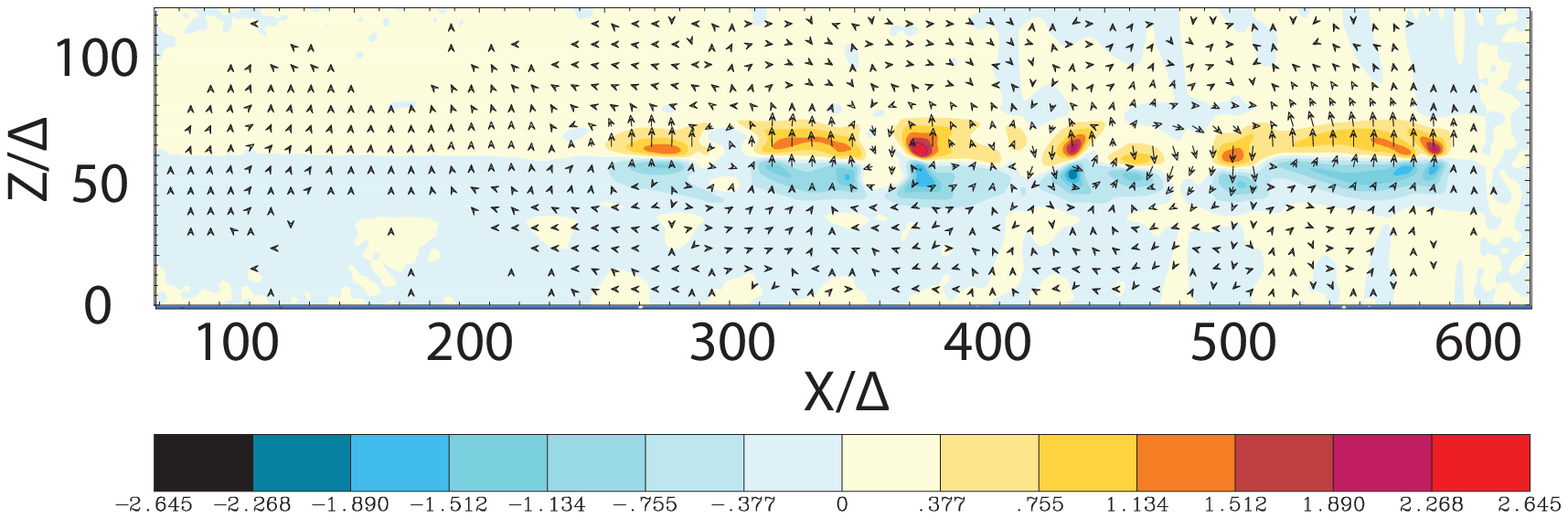}
\hspace*{-0.3cm}
\includegraphics[scale=0.43]{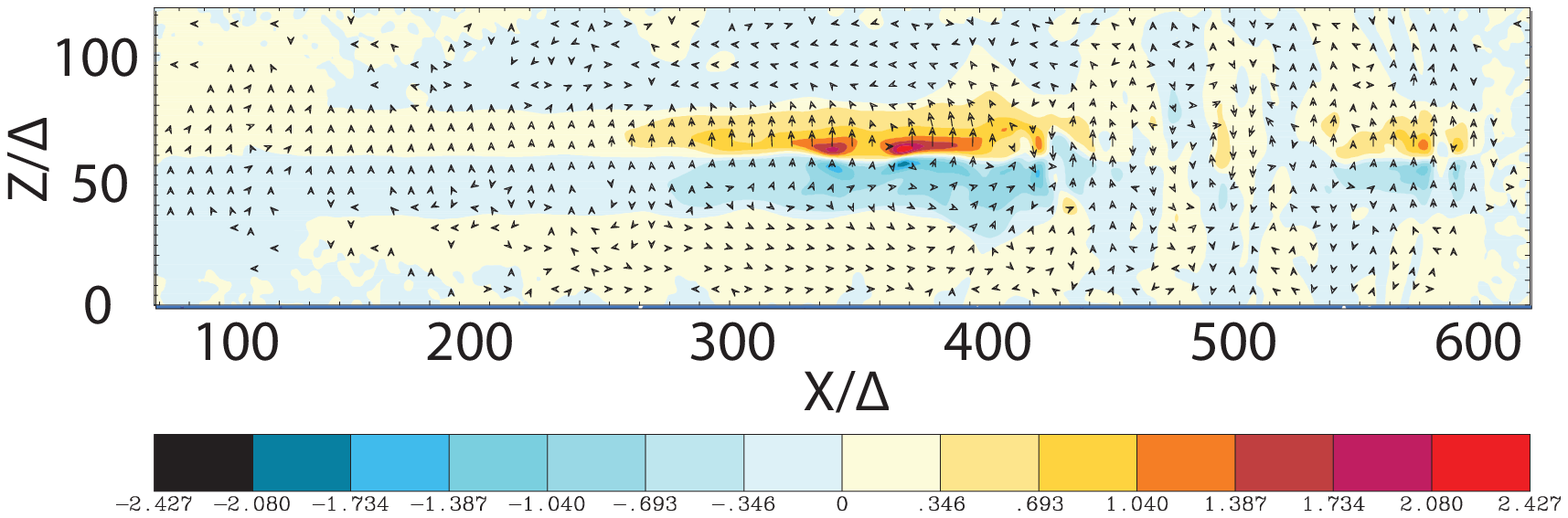}
\vspace*{-0.2cm}

\hspace{7.20cm} (c) \hspace{7.0cm} (d)

\hspace*{0.3cm}
\includegraphics[scale=0.43]{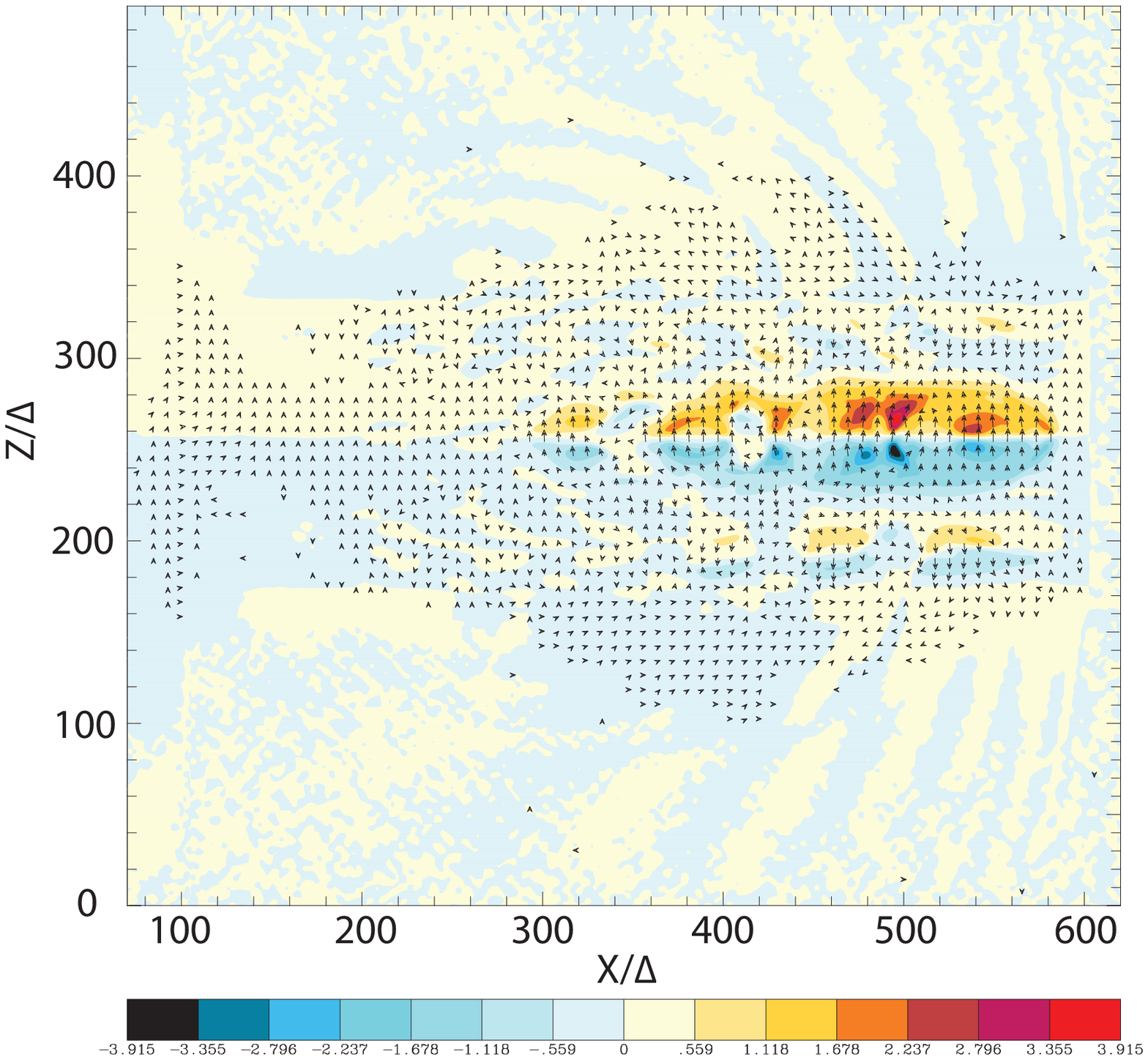}
\hspace*{-0.3cm}
\includegraphics[scale=0.43]{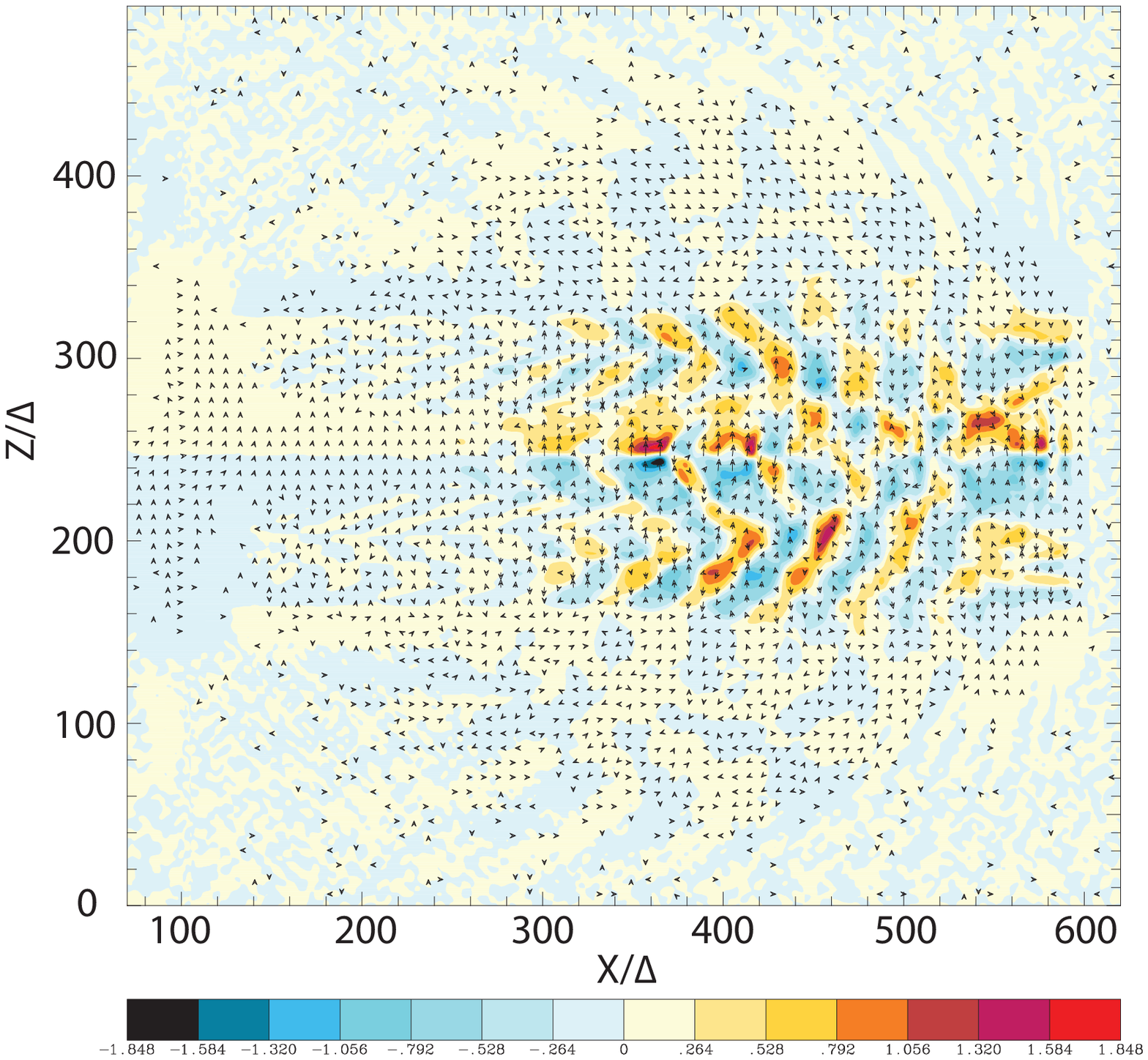}

\caption{\footnotesize{Isocontour plots of the azimuthal component of magnetic field $B_y$ intensity at the center 
of the jets  for  $e^{-}-p^{+}$  ((a) and (c))  $e^{\pm}$  ((b) and (d)) jets;  with $r_{\rm jet}=20\Delta$ ((a) and (b))  
$r_{\rm jet}=80\Delta$ ((c) and (d)) at time $t =  500\omega_{\rm pe}^{-1}$.  
The disruption of helical magnetic fields are caused by instabilities and/or  reconnection. The max/min numbers of  panels are (a) $\pm$2.645, (b) $\pm$ 2.427,
(c) $\pm$  3.915, (d) $\pm$1.848.}}
\label{ByBxz}
\vspace{-0.3cm}
\end{figure}

Figure \ref{ByBxz} shows the $y$ component of the magnetic field ($B_{y}$) in the jet radius with  $r_{\rm jet}=20\Delta\, {\rm and}\,  80\Delta$.
The initial helical magnetic field 
(left-handed; clockwise viewed from the jet front) is enhanced and disrupted due to the instabilities for both cases. 

Even as a result of shorter simulation systems,   
the growing instabilities are affected by the helical magnetic fields. These complicated patterns
of $B_{\rm y}$ are generated by the currents generated by instabilities in  jets. The larger jet radius contributes more modes of instabilities to grow in the jets, 
which  make the jet structures more complicated. The simple recollimation shock generated in the small jet radius is shown in Figs. \ref{ByBxz}(a) and 
\ref{ByBxz}(b) \cite{nishi16b}.  We need to perform
longer simulations in order to investigate full development of instabilities and jets with helical magnetic fields. 

\vspace{-0.50cm}
\begin{figure}[h!]
  
\hspace{7.3cm}(a) \hspace{7.cm}(b)

\vspace*{-0.3cm} 
\hspace*{0.3cm}
\includegraphics[scale=0.385,angle=-90]{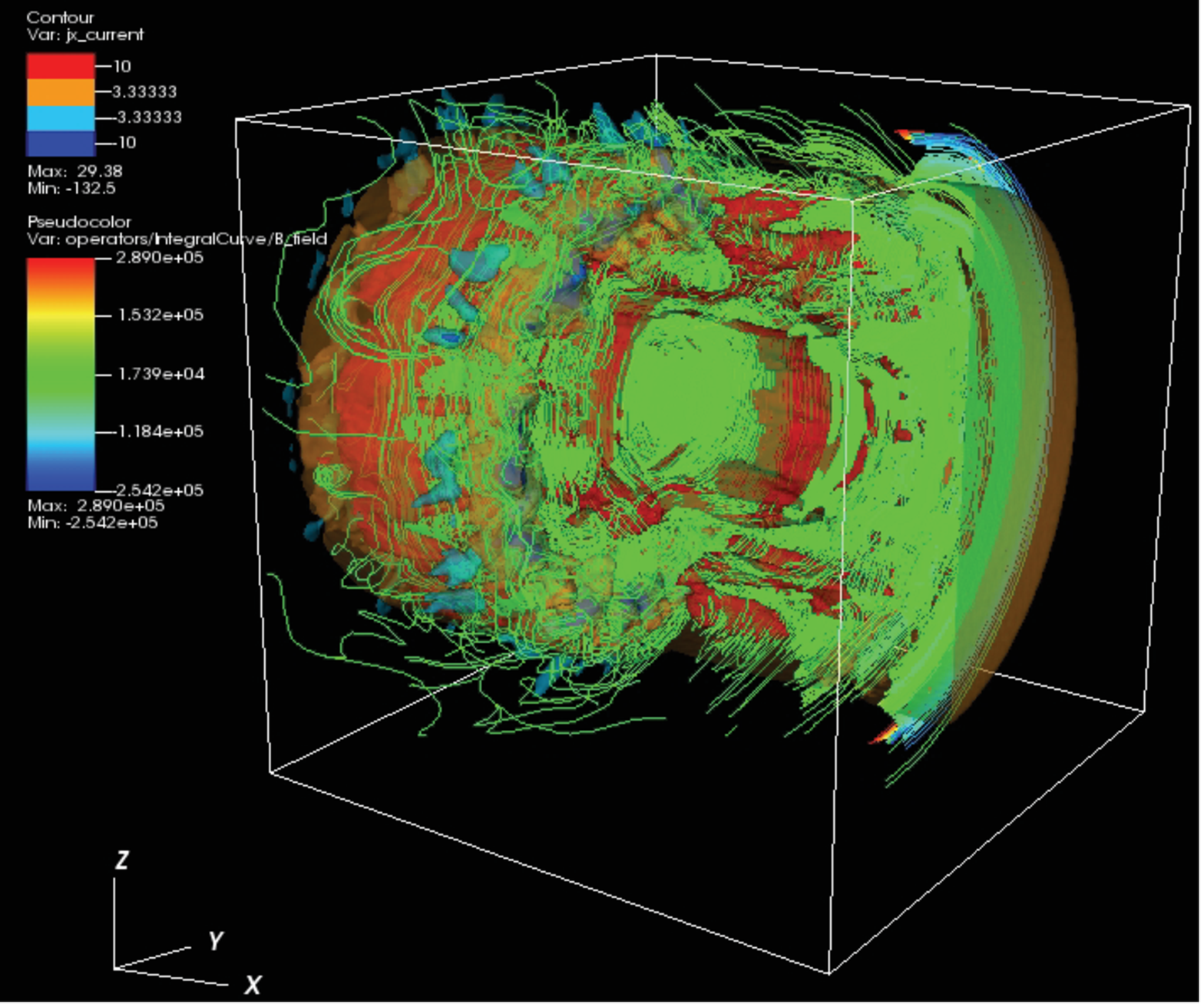}
\hspace*{-0.1cm}
\includegraphics[,scale=0.385,angle=-90]{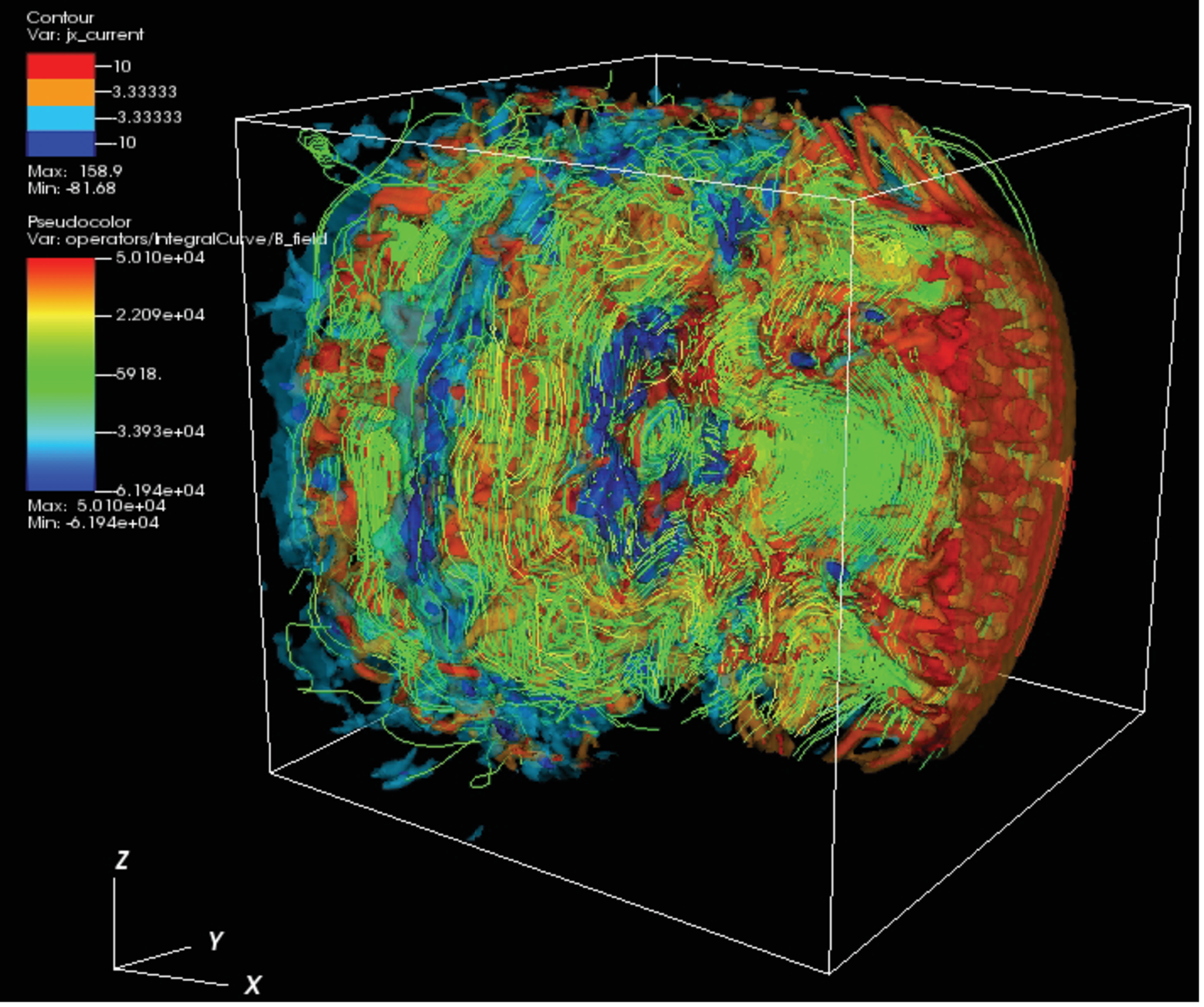}


\vspace*{-0.cm}
\caption{\footnotesize{Panels show 3D iso-surface plots of  the current ($J_{\rm x}$) of jet electrons for $e^{-}$ {--}$p^{+}$ (a) and 
$e^{\pm}$ (b) ~jet with  $r_{\rm jet}=80\Delta$  at time  $t =  500 \omega_{\rm pe}^{-1}$. The lines show the magnetic field stream lines in the quadrant
of the front part of jets. The color scales for contour (upper left) for (a) and (b): red 10; orange  3.33; right blue $-3.33$. blue $-10$.
The color scales of streaming lines (a) (2.89, 1.53, 0.174, $-1.2$, $-2.54$) $\times 10^{5}$; (b) (5.01, 2.21, $-0.592$, $-3.39$, $-6.19$) $\times 10^{4}$.}}\label{3dJxBS}\end{figure}

In order to investigate 3D structures of averaged jet electron current ($J_{\rm x}$) we plot it in the 3D ($420 \le x/\Delta \le 620$,  $152 \le y,z/\Delta \le 352$)
region of the jet front.

Figure \ref{3dJxBS} shows the current ($J_{\rm x}$) of jet electrons for $e^{-}$ {--}$p^{+}$ (a) and 
$e^{\pm}$ (b) ~jets. The cross sections at $x/\Delta = 520, y/\Delta= 252$ and surfaces of jets show complicated patterns, which are generated by instabilities
with the magnetic field lines. 


%
 

\vspace{-0.2cm}
\begin{figure}[ht]
  
\hspace{6.2cm}(a) \hspace{7.1cm}(b)

\hspace*{0.2cm}
\includegraphics[scale=0.47]{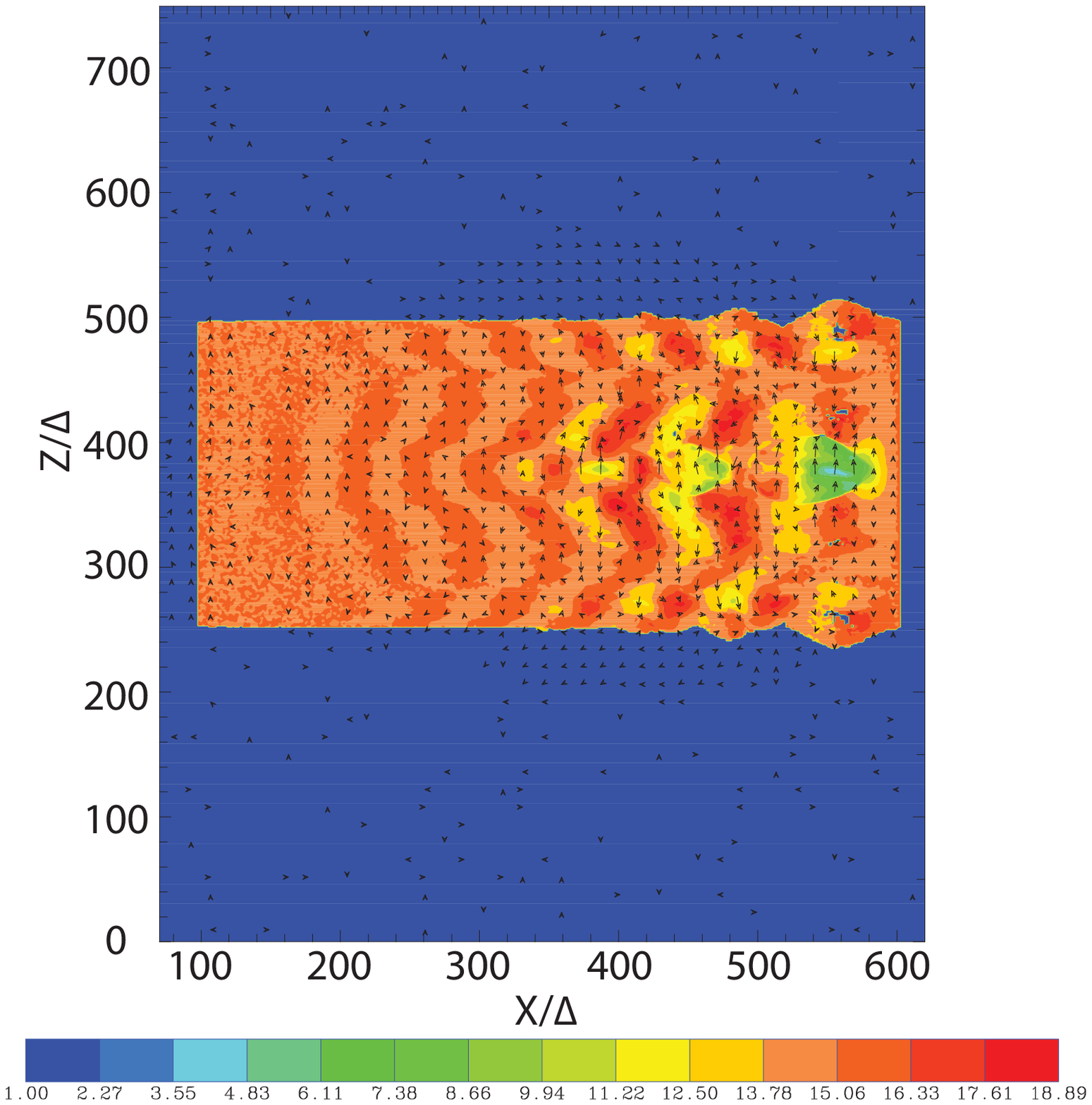}
\hspace*{-0.2cm}
\includegraphics[,scale=0.47]{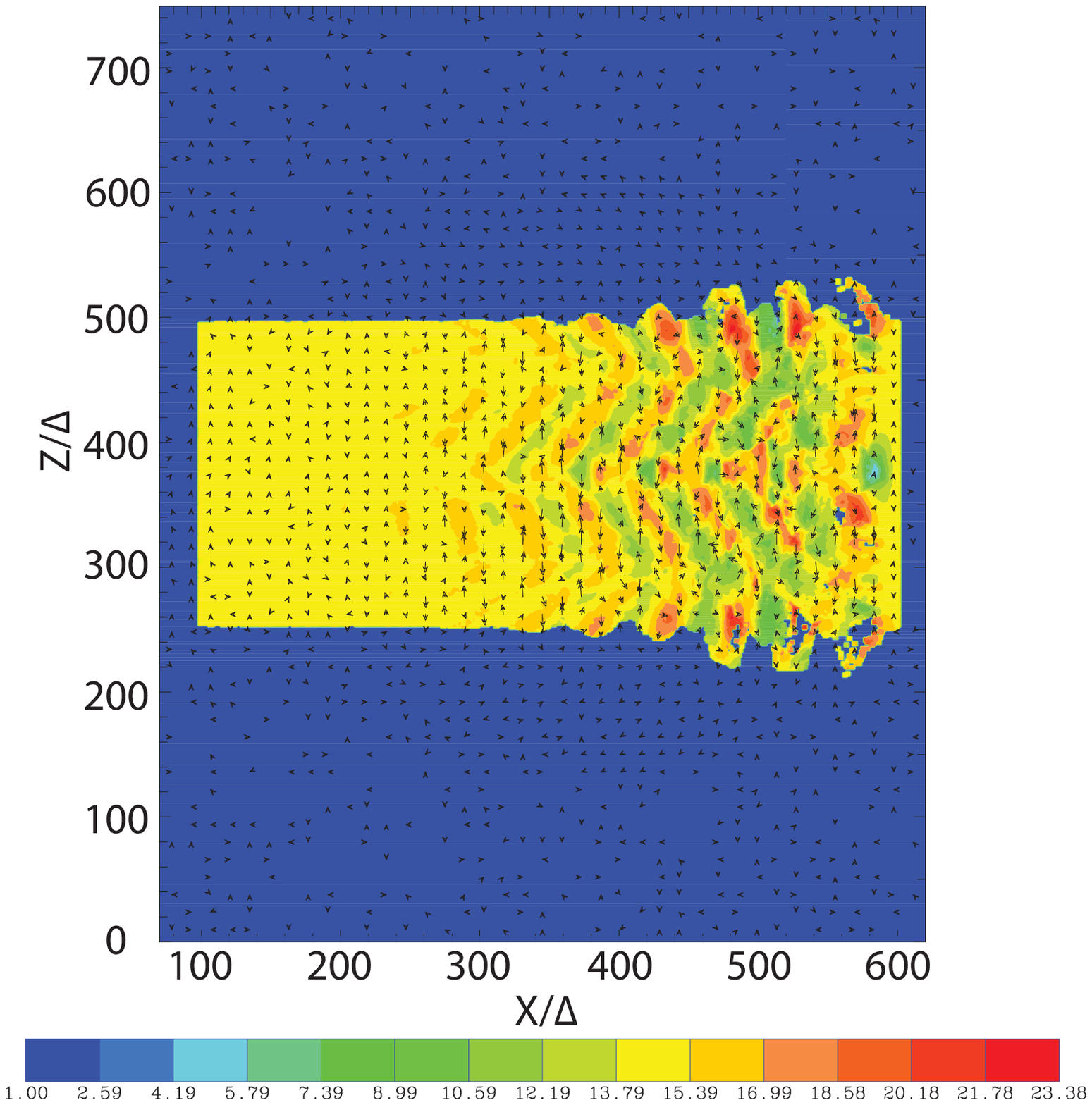}


\vspace*{-0.3cm}
\caption{\footnotesize{Panels (a) and (b) show 2D plot of  the Lorentz factor of jet electrons for $e^{-}$ {--}$p^{+}$ (a) and 
$e^{\pm}$ (b)
 ~jet with  $r_{\rm jet}=120\Delta$  at time  $t =  500 \omega_{\rm pe}^{-1}$. The arrows (black spots) show the magnetic fields in the $x-z$
 plane.}}
 \label{lor}
 \end{figure}


In order to determine particle acceleration we calculated the Lorentz factor of jet electrons in the cases with
$r_{\rm jet}=120\Delta$ as shown in Fig. \ref{lor}. These patterns of Lorentz factor  coincide with the changing directions of local magnetic fields
that are generated by instabilities.
The directions of magnetic fields are indicated by the arrows (black spots), which can be seen with magnification.
The directions of magnetic fields are determined by the generated instabilities.
The structures at the edge of jets are generated by the kKHI. 
The plots of Lorentz factor in the $y-z$ plane show the MI in the circular edge of the jets (the $x$ component of current $J_{x}$) as shown in Figure S1.

\vspace{-0.2cm}
\begin{figure}[h!]
  
\hspace{7.2cm}(a) \hspace{6.7cm}(b)  

\vspace*{-0.3cm} 

\hspace*{0.6cm}
\includegraphics[scale=0.33,angle=-90]{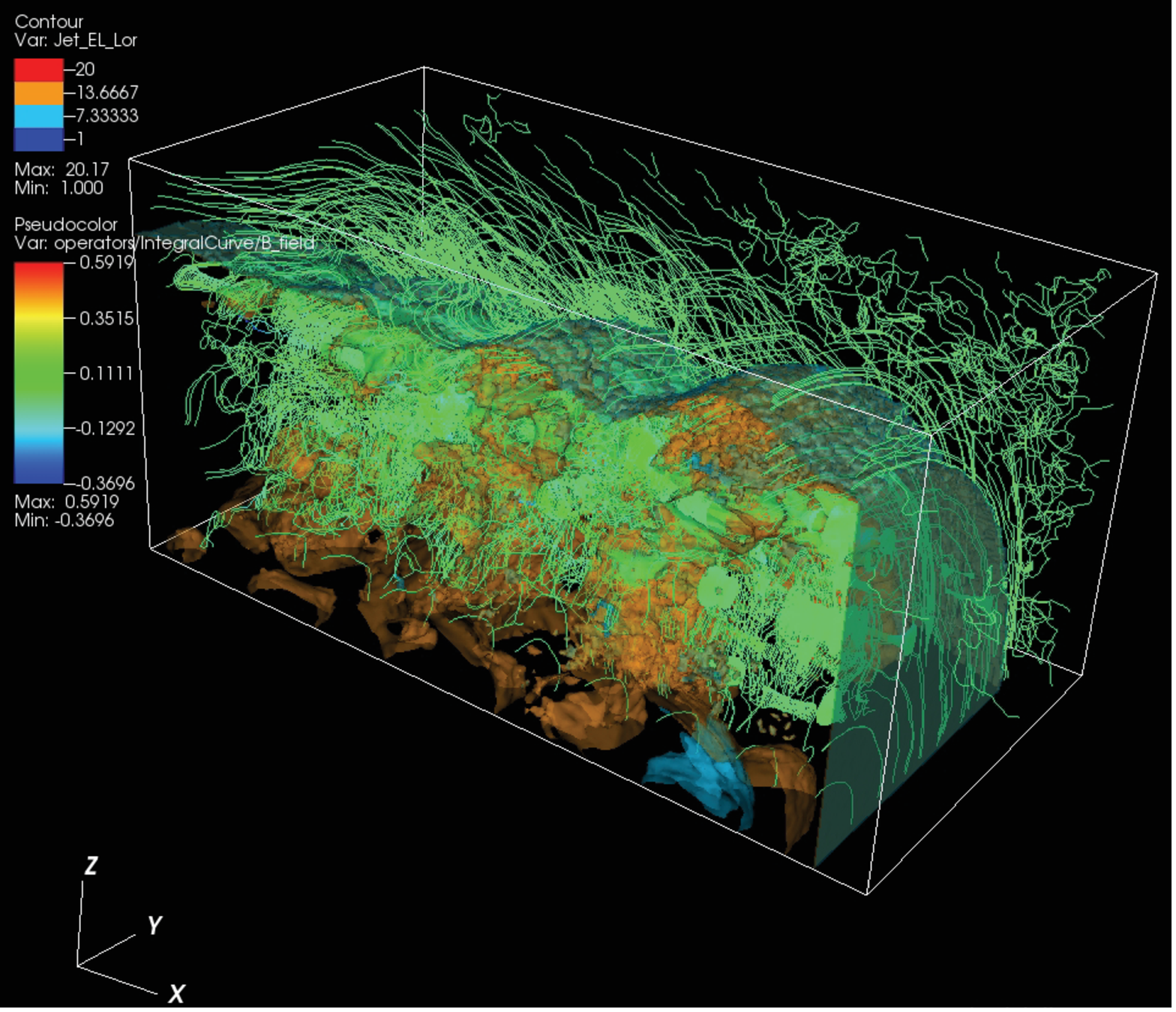}
\hspace*{-0.cm}
\includegraphics[,scale=0.33,angle=-90]{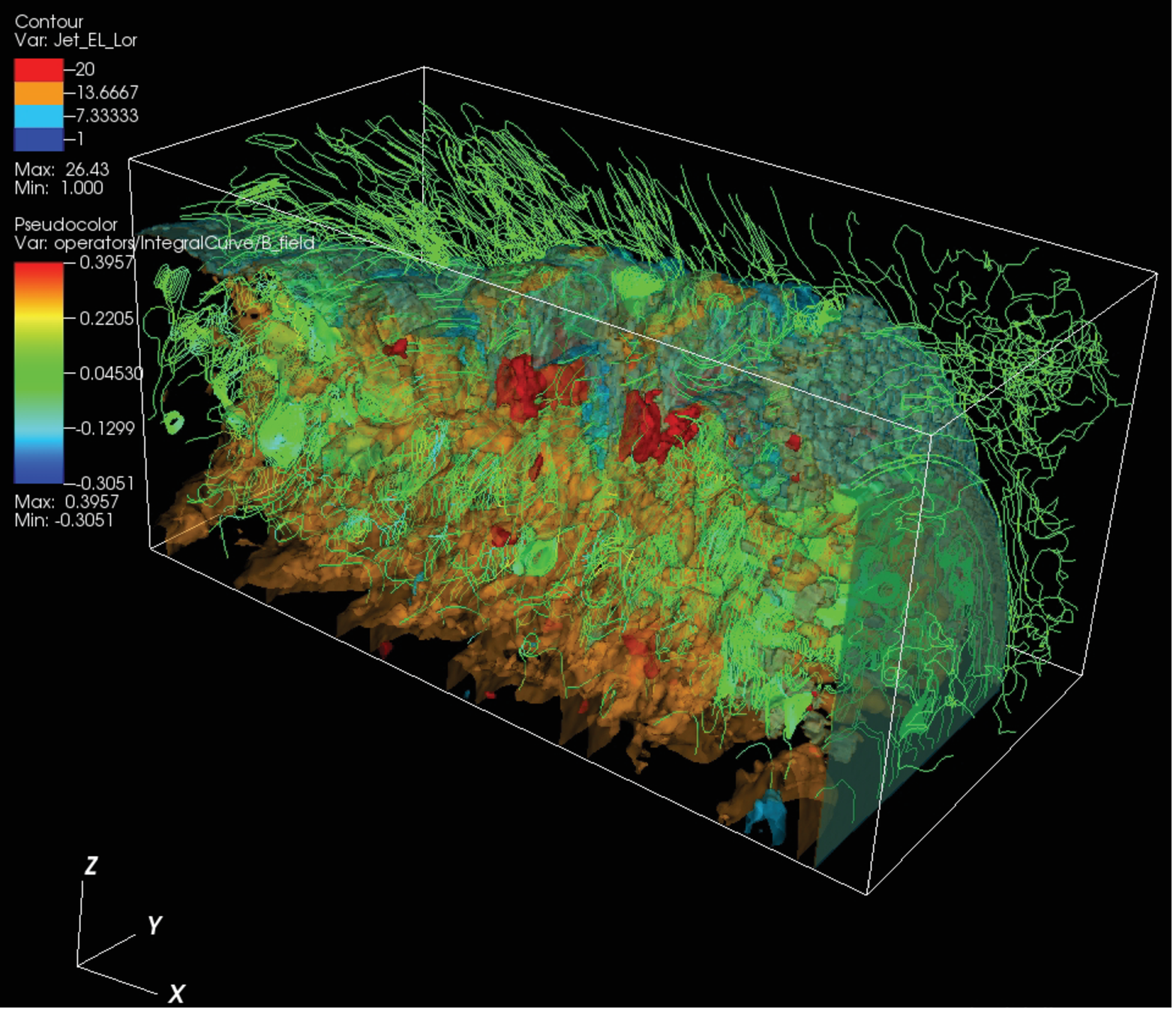}


\vspace*{-0.0cm}
\caption{\footnotesize{Panels show 3D iso-surface plots of  the Lorentz factor of jet electrons for $e^{-}$ {--}$p^{+}$ (a) and 
$e^{\pm}$ (b) ~jet with  $r_{\rm jet}=120\Delta$  at time  $t =  500 \omega_{\rm pe}^{-1}$. The lines show the magnetic field stream lines in the quadrant
of the front part of jets.
The color scales for contour (upper left) for both (a) and (b): red 20.0; orange  13.67; right blue $7.33$. blue $1$.
The color scales of streaming lines (a) (5.92, 3.52, 0.174, $-1.29$, $-3.70$) $\times 10^{-1}$; (b) (3.96, 2.21, $0.453$, $-1.30$, $-3.05$) $\times 10^{-1}$.}}\label{3dlorBS}\end{figure}

In order to investigate 3D structures of the averaged jet electron Lorentz factor, we plot iso-surface of it in 3D 
($320 \le x/\Delta \le 620$,  $381 \le y,z/\Delta \le 531$) of a quadrant of the jet front.

\vspace*{-0.cm}
Figure \ref{3dlorBS} shows the Lorentz factor of jet electrons for $e^{-}$ {--}$p^{+}$ (a) and 
$e^{\pm}$ (b) ~jet. The cross sections and surfaces of jets show complicated patterns that are generated by instabilities
with the magnetic field lines. 

In both cases with the jet radii larger than $r_{\rm jet}=80\Delta$,  at the jet surfaces kKHI and MI are generated, and inside the jets the Weibel instability 
is generated with kink-like instability, in particular in the electron-proton jet. We aim to investigate further using different parameters including $a$, 
which determines the structure of helical magnetic fields in Equations (1) and (2).

\section{Discussion}

The global jet simulations with large jet radii show the importance of a larger jet radius in RPIC simulations for investigating in-tandem the macroscopic 
processes incorporated in RMHD simulations. Due to mixed modes of generated instabilities, jet electrons in phase space show little or no bunching 
in comparison to  those with jet radius $r_{\rm jet}=20\Delta$ as shown in Figs. 5(a) and 5(b) in the previous report
\cite{nishi16b}. Consequently, recollimation shocks occur rather in the center of jets, which is dependent on
the value of $a$ in Eqs. (1) and (2). Further simulations with different  values of $a$ are required for investigating the evolution of kinetic 
instabilities in global jets. 

These simulations show that the energy stored in helical magnetic fields is released due to the excitations of kinetic instabilities such 
as kKHI, MI and the Weibel instability with kink-like instability. Consequently, electrons are accelerated and turbulent magnetic fields
are generated, which provide polarity.  

MacDonald \& Marscher \cite{mm16} have developed a radiative transfer scheme
that allows the Turbulent Extreme Multi-Zone (TEMZ) code to produce simulated images of the time-dependent linearly and
circularly polarized intensity at different radio frequencies. Using this technique with our simulation results, we have produced synthetic
polarized emission maps that highlight the linear and circular polarization expected within the model. We will discuss these findings in a separate paper.


We plan to perform additional simulations using a larger jet radius and longer systems in order to investigate the full dynamics of jet evolution
and interaction with ambient mediums via developing instabilities. 








\vspace{6pt} 

\supplementary{The following are available online at www.mdpi.com/link, Figure S1: title, Table S1: title, Video S1: title.}

\acknowledgments{This work is supported by NSF AST-0908010, AST-0908040,
NASA-NNX09AD16G, NNX12AH06G, NNX13AP-21G,
and NNX13AP14G grants. The work of J.N. and O.K. has been
supported by Narodowe Centrum Nauki through research
project DEC-2013/10/E/ST9/00662.
Y.M. is supported by the ERC Synergy Grant ``BlackHoleCam - Imaging
the Event Horizon of Black Holes''  (Grant No. 610058).  M.P.~acknowledges
support through grant PO 1508/1-2 of the Deutsche Forschungsgemeinschaft.
Simulations were performed using  Pleiades and Endeavor facilities at NASA
Advanced Supercomputing (NAS), and using Gordon and Comet at The San Diego
Supercomputer Center (SDSC), and Stampede at The Texas Advanced
Computing Center, which are supported by the NSF. This research was started during the program
``Chirps, Mergers and Explosions: The Final Moments of Coalescing Compact Binaries''
at the Kavli Institute for Theoretical Physics, which is supported by the National Science
Foundation under grant No. PHY05-51164. The first velocity shear results using
an electron$-$positron plasma were obtained during the Summer Aspen workshop
``Astrophysical Mechanisms of Particle Acceleration and Escape from the Accelerators''
held at the Aspen Center for Physics (1--15 September 2013).}

\authorcontributions{K. -I. Nishikawa: Perform simulations, analyze the data and prepare a manuscript; Y. Mizuno: Compare with RMHD simulations; J. L. G\'omez: 
Contribute for comparing simulation results to observations; J. Niemiec: Contribute modifying the code for this research; O. Kobzar; Modify the code for this simulation; M. Pohl: Overlook the simulation results; J. L. G\'{o}mez : Contribute on comparisons with observations; I. Du\c{t}an: Perform some of simulations for this research; A. Pe'er: Critical contributions for physical interpretation; J. T. Frederiksen: Contribution for critical discussions on this research; \AA. Nordlund: Fruitful suggestions for this research; C. White: Contribute some simulations and discussions; A. Meli: Critical reading and discussion on this research; H. Sol: Essential suggestions for this research; P. E. Hardee: Theoretical contributions for this research; D. H. Hartmann: Useful discussions for this research}

\conflictsofinterest{The authors declare no conflict of interest.}




\sampleavailability{Data (vtk) for VisIt and ParaView are available upon the request.}

\end{document}